\documentclass[10pt]{article}
\usepackage[cp1251]{inputenc}
\newcommand{\sss}{\scriptstyle}

\usepackage[dvips]{graphicx}   \usepackage[dvips]{epsfig}
\def\lsim{\  \lower-1.2pt\vbox{\hbox{\rlap{$<$}\lower5pt\vbox{\hbox{$\sim$}}}}\ }
\def\gsim{\  \lower-1.2pt\vbox{\hbox{\rlap{$>$}\lower5pt\vbox{\hbox{$\sim$}}}}\ }

\begin{document}
\title{Uniqueness of the solution of the Gaudin's equations,  \\ which describe
 a one-dimensional system of point bosons \\   with zero boundary conditions}
\author{ {\small Maksim Tomchenko}
\bigskip \\ {\small Bogolyubov Institute for Theoretical Physics, } \\
 {\small 14b,  Metrolohichna Str., Kyiv 03680, Ukraine} \\
 {\small E-mail: mtomchenko@bitp.kiev.ua}}
 \date{\empty}
 \maketitle
 \large
 \sloppy

\begin{abstract}
We show that the system of Gaudin's equations for quasimomenta
$k_{j}$, which describes a one-dimensional system of spinless point
bosons with zero boundary conditions, has the unique real solution
for each set of quantum numbers $n_{j}$.
\end{abstract}

 \section{Introduction}
The one-dimensional (1D) system of spinless point bosons is one of
the most investigated integrable systems
\cite{girardeau1960,ll1963,lieb1963,yangs1969,gaudin1971,gaudinm,takahashi,batchelor2005,mt2015}
(we give the basic references or those related directly to our
work). This is the simplest system. Therefore, the understanding of
its properties is of importance. The system is described by a wave
function in the form of the Bethe ansatz with quasimomenta $k_{j}$
\cite{ll1963,gaudin1971}. For the periodic boundary conditions
(BCs), it takes the form \cite{ll1963}
\begin{equation}
 \psi_{\{k \}}(x_{1},\ldots,x_{N})=\sum\limits_{P}a(P)e^{i\sum\limits_{l=1}^{N} k_{P_{l}}x_{l}},
      \label{1-1} \end{equation}
where $P$ means all permutations of $k_{l}$. The  Schr\"{o}dinger
equation and BCs lead to the equations for $k_{j}$, which contain
all information about the system: the ground-state energy,
quasiparticle dispersion law, thermodynamic quantities, {\it etc}.
For the proper determination of these quantities, it is important to
know whether the solution $\{k_{j}\}$ is unique for each collection
of quantum numbers $n_{j}$. For the periodic boundary conditions
(BCs), the uniqueness of the solution was proved by Yang and Yang
\cite{yangs1969} (more precisely, work \cite{yangs1969} gave only
the idea of a proof, and the strict proof was obtained by Takahashi
\cite{takahashi}). However, the real systems have usually
boundaries. \textit{Zero} BCs are set for the domain $0\leq
x_{1}\leq x_{2}\leq\ldots \leq x_{N}\leq L$, they read
\cite{gaudin1971,gaudinm}
\begin{equation}
 \Psi(x_{1}=0,x_{2}\ldots,x_{N})=\Psi(x_{1},\ldots,x_{N-1},x_{N}=0)=0.
      \label{1-2} \end{equation}
For such BCs, the solution should be sought in the form
\cite{gaudin1971,gaudinm}
\begin{equation}
 \Psi_{\{|k| \}}(x_{1},\ldots,x_{N})=\sum\limits_{\{\varepsilon \}}C(\varepsilon_{1},\ldots,\varepsilon_{N})\psi_{\{k
 \}}(x_{1},\ldots,x_{N}),
      \label{1-3} \end{equation}
where $\varepsilon_{j}=\pm 1$. Gaudin showed
\cite{gaudin1971,gaudinm} that the wave function (\ref{1-3})
satisfies BCs (\ref{1-2}) and the Schr\"{o}dinger equation, if the
quasimomenta $k_{j}$ satisfy the definite equations. We call them
the Gaudin's equations and give them in the next section. On the
basis of these equations, the ground state and excited levels of the
system were studied, respectively, in
\cite{gaudin1971,batchelor2005} and \cite{mt2015}. The Gaudin's
equations possess a particular symmetry, which allows us to
differently introduce quasiparticles and to construct the
thermodynamics in a new way \cite{mt2015,mt2016termod}. But the
uniqueness of the solution of the Gaudin's equations was not proved
previously. This will be made in the present work.

\section{Some properties of the Gaudin's equations}
Consider the system of $N$ spinless point bosons located on the
segment of length $L$. The  Schr\"{o}dinger equation and BCs
(\ref{1-2}) yield for this system the following equations for
quasimomenta $k_{j}$ \cite{gaudin1971}:
\begin{equation}
e^{2ik_{j}L}=\prod\limits_{l=1}^{N}\frac{(k_{j}+ic)^{2}-k_{l}^{2}}{(k_{j}-ic)^{2}-k_{l}^{2}}|_{l\neq
j}, \quad j=1,\ldots, N.
     \label{2-1} \end{equation}
Here, the product contains no factor with $l=j$, all $k_{j}$ are
nonzero real numbers, and $|k_{l}|\neq |k_{j}|$ for $l\neq j$ (the
case where $k_{j}= 0$ for one or several $j$ and the case where
$|k_{l}|= |k_{j}|$ for one or several pairs $j,l$ are excluded,
since the Schr\"{o}dinger equation with zero BCs has no solutions in
these cases). We consider only the repulsive  interaction: $c> 0$.
Equations (\ref{2-1}) are invariant relative to the changes
$k_{j}\rightarrow -k_{j}$ and (independently) $k_{l}\rightarrow
-k_{l}$. Therefore, it is sufficient to find the solutions with
$k_{j}
> 0$ for all $j$. Equations (\ref{2-1}) yield the Gaudin's equations
\cite{gaudin1971}
\begin{eqnarray}
Lk_{i}=\pi n_{i}+\sum\limits_{j=1}^{N}\left (\arctan{\frac{c}{k_{i}-k_{j}}} +
\arctan{\frac{c}{k_{i}+k_{j}}}\right )|_{j\neq i}, \ i=1,\ldots, N,
     \label{2-2} \end{eqnarray}
where $n_{i}=0, \pm 1, \pm 2, \ldots$, and $k_{i} > 0$ for all $i$.
If $c=0$, these equations possess the solution for free bosons in a
box: $Lk_{i}=\pi n_{i}$, $n_{i}= 1, 2, \ldots$ ($i=1,\ldots, N$).
Based on the idea of the continuity of solutions in the interaction,
Gaudin concluded \cite{gaudin1971} that Eqs. (\ref{2-2}) should be
solved only with positive values of $n$'s ($n_{i} = 1, 2, \ldots$),
for which there exists a continuous transition $c\rightarrow 0$ to
the solutions for free bosons. We do not agree with this argument.
Because at a fixed $c> 0$ nothing prevents the system from passing
from the state with $n_{i}> 0$ (for all $i$) in a state with
$n_{i}\leq 0$ (for one or several $i$) due to the gradual (or
jump-like) decrease of  $n$'s. Therefore, the solutions
corresponding to  $n_{i}\leq 0$ for one or several $i$ should be
taken into account. Let us clarify what are those solutions.

Let the set $\{n_{i}\}$ has  $n_{l}< 0$ for a single $l$. We
multiply the $l$th equation from (\ref{2-2}) by $-1$ and get
\begin{eqnarray}
L\tilde{k}_{l}=\pi \tilde{n}_{l}+\sum\limits_{j=1}^{N}\left (\arctan{\frac{c}{\tilde{k}_{l}-k_{j}}} +
\arctan{\frac{c}{\tilde{k}_{l}+k_{j}}}\right )|_{j\neq l},
     \label{2-3} \end{eqnarray}
where $\tilde{n}_{l}=-n_{l}>0, \tilde{k}_{l}=-k_{l}$. So, we return
to an equation of the form (\ref{2-2}), but already with a positive
$n_{l}$. This means that the solution $\{k_{i}\}$ for $n_{l}=p< 0$
differs from the solution $\{k_{i}\}$ for $n_{l}=-p>0$ only by the
sign of $k_{l}$. However, the solutions, which differ from one
another only by the sign of one or several $k_{i}$, are physically
equivalent, because they lead to the same wave function
$\Psi(x_{1},\ldots,x_{N})$ (\ref{1-3}). Therefore, it is sufficient
to find the solutions for all nonnegative $n_{i}$.

Consider the case, for which the set $\{n_{i}\}$ contains a zero
$n_{l}$. Suppose that $k_{l}=0$ in the $l$th equation in
(\ref{2-2}), then this equation is satisfied identically. We assign
the number $1$ to this equation. Then we have $k_{1}=0$, and the
remaining equations take the form
\begin{eqnarray}
Lk_{i}=\pi
n_{i}+2\arctan{\frac{c}{k_{i}}}+\sum\limits_{j=2}^{N}\left
(\arctan{\frac{c}{k_{i}-k_{j}}} +
\arctan{\frac{c}{k_{i}+k_{j}}}\right )|_{j\neq i},
     \label{2-4} \end{eqnarray}
where $i=2,\ldots,N$. In the next section, we will show that the
systems (\ref{2-2}) and (\ref{2-4}) have the unique solution. This
implies that the solution with $k_{1}=0$ and $k_{2},\ldots,k_{N}$
from (\ref{2-4}) coincides with the unique solution of system
(\ref{2-2}). If $n_{l}= 0$ for several $l$, then the solution
contains one zero $k_{i}$. However, it was noted above that all
solutions $\{k_{i}\}$ with $k_{l}=0$ for at least one $l$ should be
omitted.

Thus, the consideration of $n_{i}\leq 0$ does not lead to physically
new solutions. Therefore, in order to find all admissible
nonequivalent solutions of system (\ref{2-1}), it is sufficient to
solve system (\ref{2-2}) with $n_{i}=1,  2, \ldots$ and  $k_{i} > 0$
for all $i$. This coincides with the conclusion made by Gaudin
\cite{gaudin1971,gaudinm}.

  \section{Uniqueness of the solution of the Gaudin's equations}
We will prove the uniqueness of the solution of the Gaudin's
equations (\ref{2-2}) with the help of the  methods by Yang and Yang
\cite{yangs1969} and by Takahashi \cite{takahashi}.

First, we recall the Gaudin's idea \cite{gaudin1971,gaudinm} of that
the equations for a periodic system \cite{gaudin1971}
\begin{eqnarray}
Lk_{i}= 2\pi
n_{i}+2\sum\limits_{j=1}^{N}\arctan{\frac{c}{k_{i}-k_{j}}}|_{j\neq
i}, \quad i=1,\ldots, N,             \label{3-1} \end{eqnarray} can
be written in the form (\ref{2-2}). Let $N$ be even, and let the set
of quantum numbers $n_{i}$ is specularly antisymmetric:
$n_{l}=-n_{N-l+1}$, $l=1,\ldots,N/2$. Then Eqs. (\ref{3-1}) can be
written in the form (\ref{2-2}) with $N\rightarrow N/2$,
$L\rightarrow L/2,$ and the additional term $\arctan{(c/2k_{i})}$ on
the right-hand side (see also Eq. (21) in \cite{mt2015}). As a
result, the equations almost coincide with (\ref{2-2}). But we have
no complete coincidence due to the term $\arctan{(c/2k_{i})}$.
Therefore, the solutions for zero BCs do not belong to the solution
set for periodic BCs. Hence, the uniqueness of solutions with zero
BCs should be proved separately. This problem is more complicated
than that for periodic BCs.

As was noted above, it is sufficient to solve Eqs. (\ref{2-2}) in
the case where all $k_{j}$ are different and positive. Therefore, we
may always order $k$'s so that
\begin{equation}
0<k_{1}<k_{2}<\ldots < k_{N}. \label{3-21} \end{equation}
It follows from (\ref{2-2}) that
\begin{eqnarray}
&&L(k_{i+1}-k_{i})=\pi
(n_{i+1}-n_{i})+2\arctan{\frac{c}{k_{i+1}-k_{i}}}\label{3-22}
\\&+&\sum\limits_{j=1}^{i-1}\left
(\arctan{\frac{c}{k_{i+1}-k_{j}}}-\arctan{\frac{c}{k_{i}-k_{j}}}\right
) + \sum\limits_{j=1}^{i-1}\left (\arctan{\frac{c}{k_{i+1}+k_{j}}}-
\arctan{\frac{c}{k_{i}+k_{j}}}\right )\nonumber \\ &+&
\sum\limits_{j=i+2}^{N}\left (\arctan{\frac{c}{k_{i+1}-k_{j}}}
-\arctan{\frac{c}{k_{i}-k_{j}}}\right )+
\sum\limits_{j=i+2}^{N}\left (\arctan{\frac{c}{k_{i+1}+k_{j}}}
-\arctan{\frac{c}{k_{i}+k_{j}}}\right ).
   \nonumber  \end{eqnarray}
At the ordering  (\ref{3-21}), each of four sums on the right-hand
side of (\ref{3-22}) is negative. Therefore, in order that
$k_{i+1}-k_{i}> 0,$ it is necessary that $\pi
(n_{i+1}-n_{i})+2\arctan{\frac{c}{k_{i+1}-k_{i}}}>0$. From whence,
we have $n_{i+1}-n_{i}\geq 0$. That is, the ordering (\ref{3-21}) is
possible only if
\begin{equation}
n_{1}\leq n_{2}\leq \ldots \leq n_{N}. \label{3-23} \end{equation}
Consider firstly the case where all $n_{i}$ are different:
\begin{equation}
n_{1}< n_{2}< \ldots < n_{N}. \label{3-24} \end{equation} Using the
relation
\begin{equation}
\arctan{\alpha}=(\pi/2)sgn(\alpha)-\arctan{(1/\alpha)} \label{3-a}
\end{equation}
and ordering (\ref{3-21}), we now pass from (\ref{2-2}) to the
equivalent equations
\begin{eqnarray}
Lk_{i}=\pi I_{i}-\sum\limits_{j=1}^{N}\left
(\arctan{\frac{k_{i}-k_{j}}{c}}+
\arctan{\frac{k_{i}+k_{j}}{c}}\right )\left. \right |_{j\neq i},
     \label{3-25} \end{eqnarray}
\begin{equation}
I_{i}=n_{i}+i-1.
     \label{3-26} \end{equation}
Since Eq. (\ref{3-a}) is not defined for $\alpha=0$, systems
(\ref{3-25}) and (\ref{2-2}) are not equivalent for such collections
$\{k_{i}\}$, for which $k_{i}+ k_{j}=0$ or $k_{i}- k_{j}=0$ for some
$i\neq j$. But this is insignificant. Since system (\ref{2-2}) is
not defined for $k_{i}\pm k_{j}=0$,  and system (\ref{3-25}) has no
solutions for $k_{i}\pm k_{j}=0$. Indeed, relation $k_{i}+ k_{j}=0$
requires $I_{i}+I_{j}=0$, which means $n_{i}+i+n_{j}+j=2$. But this
is impossible in view of $i, j, n_{i}, n_{j} \geq 1$. The condition
$k_{i}- k_{j}=0$ requires $I_{i}-I_{j}=0$, i.e., $n_{i}+i=n_{j}+j$.
This is also impossible, since, for $i<j,$ ordering (\ref{3-24})
requires $n_{i}<n_{j}$, and, for $i>j,$ the inequality $n_{i}>n_{j}$
should hold. Therefore, we may assert that (i) systems (\ref{3-25})
and (\ref{2-2}) are equivalent in the whole domain of existence of
solutions, and (ii) their solutions coincide.

Following the Yang and Yang's idea \cite{yangs1969}, we introduce
the function
\begin{eqnarray}
B\{k\}=\sum\limits_{j=1}^{N}\left (\frac{Lk_{j}^{2}}{2}-\pi
I_{j}k_{j}\right )+ \frac{1}{2}\sum\limits_{j,l=1}^{N \prime}\left
(\int\limits_{0}^{k_{j}-k_{l}}\arctan{(k/c)} d k+
\int\limits_{0}^{k_{j}+k_{l}}\arctan{(k/c)} d k \right ),
     \label{3-2} \end{eqnarray}
where the prime over the sum means $j \neq l$. On the whole interval
of integration, the integrands are continuous. Hence, the integral
can be differentiated with respect to the upper limit. The points of
extremum of the function $B(k_{1},k_{2},\ldots,k_{N})$ are set by
the equations
\begin{equation}
\frac{\partial B}{\partial k_{i}}=0, \quad i=1,\ldots, N,
     \label{3-3} \end{equation}
which coincide with (\ref{3-25}). The matrix of the second derivatives reads
\begin{eqnarray}
B_{ij}=\frac{\partial^{2}B}{\partial k_{i}\partial
k_{j}}&=&\delta_{ij}\left (L-
\frac{2c}{c^{2}+4k_{i}^{2}}+\sum\limits_{l=1}^{N}\frac{c}{c^{2}+(k_{i}-k_{l})^{2}}+
\sum\limits_{l=1}^{N}\frac{c}{c^{2}+(k_{i}+k_{l})^{2}} \right )+ \nonumber \\
&+&\frac{c}{c^{2}+(k_{i}+k_{j})^{2}}-\frac{c}{c^{2}+(k_{i}-k_{j})^{2}}=B_{ji},
      \label{3-4} \end{eqnarray} where $i,j=1,\ldots,N$.
This matrix is positive definite, since, for a set of any real
numbers $u_{j}$ and $k_{j}$, the following relation holds at $c>0$:
\begin{eqnarray}
\sum\limits_{i,j=1}^{N }u_{i}B_{ij}u_{j}=\sum\limits_{j=1}^{N
}u_{j}^{2}L+\sum\limits_{j,l}^{j<l}\left
[\frac{c(u_{j}-u_{l})^{2}}{c^{2}+(k_{j}-k_{l})^{2}}+\frac{c(u_{j}+u_{l})^{2}}{c^{2}+(k_{j}+k_{l})^{2}}\right
]\geq 0.
      \label{3-5} \end{eqnarray}
Moreover, the right-hand side of (\ref{3-5}) is zero only if
$u_{j}=0$ for all $j$.

Or otherwise, the necessary and sufficient condition of positive
definiteness of a real symmetric matrix $B_{ij}$ (with
$i,j=1,\ldots,N$) is as follows \cite{horn1986}:
\begin{equation}
G_{j}> 0, \quad j=1,\ldots, N,
     \label{3-6} \end{equation}
where  $G_{j}$ are the dominant minors:
\begin{equation}
G_{j}=det(B_{il}), \quad i,l=1,\ldots, j.
     \label{3-7} \end{equation}
In particular, $G_{1}=B_{11}$, $G_{2}=B_{11}B_{22}-B_{12}B_{21}$,
and so on,  $G_{N}$ is the determinant of the entire  matrix
$B_{ij}$. We determined numerically minors $G_{j}$ for various sets
of $\{k_{i}\}$ for $N=10, 100,$ and $1000$. We have studied  the
homogeneous distributions of $\{k_{i}\}$ and the inhomogeneous ones,
with a small step $\{{\sss{\triangle}} k_{i}\}$ at the transition to
the following configuration. For all sets of $\{k_{i}\},$ we have
found
\begin{equation}
0<G_{1}<G_{2}<\ldots < G_{N}.
     \label{3-8} \end{equation}
This ensure the validity of criterion (\ref{3-6}).

Let us now forget condition (\ref{3-21}) and  consider that
$k_{j}\in [-\infty,\infty]$ for all $j$. In this case, the matrix of
the second derivatives of the function $B(k_{1},k_{2},\ldots,k_{N})$
(\ref{3-2}) is positive definite, and we have $B\rightarrow +\infty$
as $k_{i}\rightarrow \pm \infty$ for all $i$. Such function
$B(k_{1},k_{2},\ldots,k_{N})$ must have only one stationary point
(i.e., the point, at which relation (\ref{3-3}) is satisfied),
namely, a minimum. The second stationary point cannot exist: A
maximum and a saddle point are excluded, because the matrix $B_{ij}$
must not be positive definite in vicinities of these points. But,
according to (\ref{3-5}), $B_{ij}$ is positive definite for any real
$k$'s. The second minimum is impossible too, since at least one
stationary point would exist between two minima; and, in a vicinity
of this point, the matrix $B_{ij}$ must not be positive definite. We
conclude that the function $B(k_{1},\ldots,k_{N})$ has one and only
one stationary point, and it is a minimum. Therefore, the system of
equations (\ref{3-3}) (and, hence, systems (\ref{3-25}) and
(\ref{2-2})) has one and only one real solution $\{k_{i}\}$.

Such conclusion is valid under condition (\ref{3-24}), according to
which all $n_{j}$ are different. Let the set $\{n_{j}\}$ has several
identical $n_{j}$. In this case, the equations in system (\ref{2-2})
corresponding to identical $n_{j}$ are indistinguishable. This means
that the  numbers $k_{j}$ corresponding to identical numbers $n_{j}$
can be interchanged in the solution $\{k_{i}\}$. All such sets
$\{k_{i}\}$ are solutions of system (\ref{2-2}). The ordering
(\ref{3-21}) is proper only for one set $\{k_{i}\}$. For the
remaining sets, Eq. (\ref{2-2}) leads to Eq. (\ref{3-25}) with other
sets of numbers $I_{i}$. However, all these solutions differ from
one another only by a permutation of quasimomenta $k_{j}$ and,
therefore, are physically equivalent. We consider them as one
solution.

Thus, we have proved that, for any set of real numbers $n_{i},$ the
system of Gaudin's equations (\ref{2-2}) has one and only one real
solution $\{k_{i}\}$. Our analysis does not prove that the solutions
for certain $n$'s satisfy condition (\ref{3-21}). But the direct
numerical solution of system (\ref{2-2}) shows that, if the
inequality $0< n_{1}\leq n_{2}\leq \ldots \leq n_{N}$ holds, the
solutions satisfy condition (\ref{3-21}).

If we would construct $B(k_{1},\ldots,k_{N})$ on the basis of Eq.
(\ref{2-2}) instead of (\ref{3-25}), then the first derivatives of
$B(k_{1},\ldots,k_{N})$ would vary by a jump on the surfaces
$k_{i}\pm k_{j}=0$. These surfaces divide the manifold $\{k_{i}\}$
into a huge number of domains. In this case, the matrix $B_{ij}$
inside each domain is defined by formula (\ref{3-4}) and is positive
definite. But, due to a jump of the functions $\partial B/\partial
k_{j}$ on the boundaries of domains, the function
$B(k_{1},\ldots,k_{N})$ can have \textit{many} minima (at most one
in each domain). In such approach, it is difficult to prove the
uniqueness of solutions in the whole domain $k_{j}\in
[-\infty,\infty]$ (for all $j$). The transition from system
(\ref{2-2}) to system (\ref{3-25}) allows us to avoid this
difficulty.

For Section 2, we need to prove also the uniqueness  of the solution
of system (\ref{2-4}). The proof can be carried on like that for
system (\ref{2-2}). For the collection $0<k_{2}<\ldots < k_{N},$ we
pass from (\ref{2-4}) to the equations
\begin{eqnarray}
Lk_{i}=\pi
I_{i}-2\arctan{\frac{k_{i}}{c}}-\sum\limits_{j=2}^{N}\left
(\arctan{\frac{k_{i}-k_{j}}{c}}+
\arctan{\frac{k_{i}+k_{j}}{c}}\right )\left. \right |_{j\neq i}
     \label{3-90} \end{eqnarray}
with $i=2,\ldots,N$ and $I_{i}=n_{i}+i-1$. For system
(\ref{3-90}), we find
\begin{eqnarray}
B\{k\}=\sum\limits_{j=2}^{N}\left (\frac{Lk_{j}^{2}}{2}-\pi
I_{j}k_{j}+2\int\limits_{0}^{k_{j}}\arctan{\frac{k}{c}} d k \right
)+ \frac{1}{2}\sum\limits_{j,l=2}^{N \prime}\left
(\int\limits_{0}^{k_{j}-k_{l}}\arctan{\frac{k}{c}} d k+
\int\limits_{0}^{k_{j}+k_{l}}\arctan{\frac{k}{c}} d k \right ),
     \label{3-9} \end{eqnarray}
\begin{eqnarray}
B_{ij}&=&\delta_{ij}\left (L-
\frac{2c}{c^{2}+4k_{i}^{2}}+\frac{2c}{c^{2}+k_{i}^{2}}+\sum\limits_{l=2}^{N}\frac{c}{c^{2}+(k_{i}-k_{l})^{2}}+
\sum\limits_{l=2}^{N}\frac{c}{c^{2}+(k_{i}+k_{l})^{2}} \right )+ \nonumber \\
&+&\frac{c}{c^{2}+(k_{i}+k_{j})^{2}}-\frac{c}{c^{2}+(k_{i}-k_{j})^{2}}=B_{ji},
\quad i,j=2,\ldots,N,
      \label{3-10} \end{eqnarray}
\begin{eqnarray}
\sum\limits_{i,j=2}^{N }u_{i}B_{ij}u_{j}=\sum\limits_{j=2}^{N
}u_{j}^{2}\left (L+\frac{2c}{k_{j}^{2}+c^{2}}\right
)+\sum\limits_{j,l}^{j<l}\left
[\frac{c(u_{j}-u_{l})^{2}}{c^{2}+(k_{j}-k_{l})^{2}}+\frac{c(u_{j}+u_{l})^{2}}{c^{2}+(k_{j}+k_{l})^{2}}\right
]\geq 0,
      \label{3-11} \end{eqnarray}
where $\{k\}=k_{2},k_{3},\ldots,k_{N}$, and the prime above the sum
in (\ref{3-9}) means $j \neq l$. Similarly to the above analysis, we
extend the admissible domain for each $k_{j}$ to $
[-\infty,\infty]$. Then, in view of formulae
(\ref{3-90})--(\ref{3-11}), we conclude that system (\ref{2-4}) has
the unique real solution $(k_{2},\ldots,k_{N})$. It is possible to
verify dy direct numerical solution of Eqs. (\ref{2-4}) that,
provided $0< n_{2}< n_{3}< \ldots < n_{N},$ the inequalities $0<
k_{2}< k_{3}< \ldots < k_{N}$ are valid. If several $n_{i}$ in
(\ref{2-4}) coincide, we have several physically equivalent
solutions, which differ only by a permutation of quasimomenta
$k_{i}$. As above, we consider them as one solution.

 \section{Conclusion}
We have shown that, in order to get all solutions of  Gaudin's
equations (\ref{2-2}), it is sufficient to consider the quantum
numbers $n_{j}=1,2,3,\ldots$ for all $j$. We have also proved that,
for any set of different real numbers $n_{j},$ the system of
equations (\ref{2-2}) has the unique real solution $\{k_{j}\}$. If
the numbers $n_{j}$ include $p$ identical ones ($n_{l+1}=\ldots
=n_{l+p}$), then the Gaudin's system (\ref{2-2}) has $p!$ physically
equivalent solutions $\{k_{j}\}$, which differ only by a permutation
of quasimomenta $k_{l+1},\ldots,k_{l+p}$.

   \section*{Acknowledgments}

The author thanks Yu. Bespalov for the discussion.

     \renewcommand\refname{}

       \end{document}